\documentclass[12pt,english,review]{elsarticle}
\usepackage{mathpazo}

\usepackage[T1]{fontenc}
\usepackage[latin9]{inputenc}
\usepackage{geometry}
\usepackage{xcolor}
\usepackage{float}
\usepackage{units}
\usepackage{bm}
\usepackage{amsmath}
\usepackage{amssymb}
\usepackage{cancel}
\usepackage{graphicx}
\usepackage{wasysym}

\makeatletter

\newcommand{\noun}[1]{\textsc{#1}}
\providecommand{\tabularnewline}{\\}
\floatstyle{ruled}
\newfloat{algorithm}{tbp}{loa}
\providecommand{\algorithmname}{Algorithm}
\floatname{algorithm}{\protect\algorithmname}

\usepackage[version=3]{mhchem}
\usepackage{amsmath}
\usepackage{algorithm,algpseudocode} 
\usepackage{xcolor}
\usepackage{hyperref}

\makeatother

\usepackage{babel}
\begin{document}

\title{An implicit, conservative and asymptotic-preserving electrostatic
particle-in-cell algorithm for arbitrarily magnetized plasmas in uniform
magnetic fields}

\author[]{G. Chen\corref{cor1}}

\ead{gchen@lanl.gov}

\author[]{L. Chac\'{o}n}

\cortext[cor1]{Corresponding author}

\address{Los Alamos National Laboratory, Los Alamos, NM 87545}
\begin{abstract}
We introduce a new electrostatic particle-in-cell algorithm capable
of using large timesteps compared to particle gyro-period under a
uniform external magnetic field. The algorithm extends earlier electrostatic
fully implicit PIC implementations with a new asymptotic-preserving
particle-push scheme that allows timesteps much larger than particle
gyroperiods. In the large-timestep limit, the integrator preserves
all particle drifts, while recovering the full orbit for small timesteps.
The scheme allows for a seamless, efficient treatment of particles
with coexisting magnetized and unmagnetized species, and conserves
energy and charge exactly without spoiling implicit solver performance.
We demonstrate by numerical experiment with several problems of variable
species magnetization (diocotron instability, modified two-stream
instability, and drift instability) that orders of magnitude wall-clock-time
speedups vs. the standard fully implicit electrostatic PIC algorithm
are possible without sacrificing solution accuracy.
\end{abstract}
\maketitle

\section{Introduction}

Implicit particle-in-cell (PIC) algorithms (see e.g. Refs. \citep{chen2011energy,lapenta2011particle,chen2014energy,chen2015multi,lapenta2017exactly}
and references therein for recent implementations) may not be limited
by the stability CFL timestep constraints of explicit PIC algorithms,
and therefore hold much potential for more efficient kinetic plasma
simulations. The fully implicit nonlinear PIC variety \citep{chen2011energy,chen2014energy,chen2015multi}
is unconditionally stable, achieves simultaneous energy and charge
conservation, ensuring long-term simulation fidelity by employing
particle sub-cycling (a multirate integration approach), and enforcing
nonlinear convergence of the particle-field system. The efficiency
gains of implicit over explicit PIC reported in the literature \citep{chen2011energy,lapenta2011particle,chen2014energy,chen2015multi}
originates largely from using large spatial cells (assuming constant
number of particles per cell), but not from their ability to use large
timesteps, due to the particle subcycling strategy needed to capture
orbits accurately. Moreover, simultaneous energy and charge conservation
demanded that particles do not step over cell faces in a single substep
\citep{chen2011energy}, requiring frequent orbit interruptions and
nonlinear convergence for the orbit equations at each substep. The
cost of particle subcycling in fully implicit PIC algorithms is particularly
vexing in the context of strongly magnetized plasmas, where large
gyrofrequencies will force commensurately small timesteps for accuracy.
To make these methods competitive, an orbit integration strategy that
can accurately step over the gyroperiod is needed.

Large timesteps compared to the gyroperiod (i.e., $\omega_{c}\Delta t\apprge1$)
can be used for the integration of the particle orbit equations without
accuracy loss under certain conditions. This has been demonstrated
for the explicit Boris particle-orbit integrator (``pusher'' for
short) \citep{parker1991numerical} as well as implicit ones \citep{barnes1983implicit,vu1995accurate}.
These approaches capture most (if not all) zeroth- and first-order
drifts of the guiding-center motion accurately, including the $\mathbf{E}\times\mathbf{B}$
drift, polarization drift, etc. However, significant drawbacks remain.
In the explicit pusher, the gyroradius $\rho_{c}$ scales with $\Delta t$
(i.e., $\rho_{c}\sim O(v_{\perp}\Delta t)$, with $v_{\perp}$ the
gyro-speed), which leads to significant orbit errors in $\rho_{c}$
when $\omega_{c}\Delta t\apprge2$ \citep{parker1991numerical}. Consequently,
in practice one must limit the timestep to a fraction of the gyro-period
for accuracy \citep{levy1968computer}. De-centered implicit pushers
\citep{barnes1983implicit} introduce numerical damping that not only
reduces the gyro-radius in time, but also breaks energy conservation,
and are therefore unsuitable for long-term simulations. Time-centered
Crank-Nicolson (CN) orbit integration schemes, however, show much
promise. The CN scheme preserves the gyro-radius exactly, enables
exact energy conservation, and retains the drift motions that persist
at zero gyro-radius except for $\nabla B$ drifts, which require additional
terms. In Ref. \citep{vu1995accurate}, a $\nabla B$ force was added
to a CN integrator to account for the $\nabla B$ drift, but at the
expense of energy conservation, allowing the magnetic field to do
work on particles. More recent work \citep{ricketson2020energy} has
shown that $\nabla B$-drift terms can be added to CN in a energy-conserving
manner. We refer the reader to Ref. \citep{ricketson2020energy} for
a discussion of the method, and for a more detailed accuracy comparison
between various pushers.

\textcolor{black}{There has been much renewed interest in PIC schemes
for strong magnetization regimes, and in particular in the development
of asymptotic preserving (AP) \citep{filbet2016asymptotically,filbet2017asymptotically,filbet2018vlasov,filbet2021convergence}
and uniformly accurate (UA) \citep{frenod2015long,crouseilles2017uniformly,chartier2020uniformly}
methods. While AP schemes ensure numerical convergence to the asymptotic
solution in the strongly magnetized regime, UA schemes seek numerical
accuracy that is independent of the value of the magnetization, which
is a stronger requirement. However, to date, UA schemes are restricted
to either uniform magnetic fields \citep{frenod2015long,crouseilles2017uniformly},
or spatially varying but with uniform magnitude \citep{chartier2020uniformly}
(which removes the impact of the $\nabla B$ force on the long-time
behavior of the system). Of particular relevance to this study is
the approach proposed in Ref. \citep{filbet2017asymptotically}, where
an AP PIC scheme is constructed with an additional force term in the
particle orbit equations, very similar to that proposed in Ref. \citep{vu1995accurate},
but again without strict conservation of energy.}

In this study, we demonstrate the feasibility of an AP \citep{jin1999efficient}
implicit PIC algorithm that can employ particle substeps orders of
magnitude larger than the gyroperiod without accuracy degradation
or loss of strict conservation properties, and as a result is orders-of-magnitude
faster than the subcycled fully implicit PIC variety for strongly
magnetized plasmas \citep{chen2011energy}. For simplicity, we consider
electrostatic collisionless plasmas in the presence of uniform magnetic
fields. For these plasmas, CN captures \emph{all} relevant drifts,
and therefore can provide the basis for the present study. When implemented
with a novel cell-crossing scheme that is both efficient and energy-
and charge-conserving (similar in concept to that proposed in \citep{chen2020semi}
for Vlasov-Maxwell PIC), the AP CN scheme proposed here delivers accurate
and efficient kinetic simulations using large timesteps (i.e., $\omega_{ce}\Delta t\gg1$)
but still compatible with dynamical time scales of interest. Moreover,
since the numerical gyro-frequency ($\hat{\omega}_{ce}$) decreases
with the timestep according to the solution of $\hat{\omega}_{ce}\Delta t/2=\mathrm{atan}(\omega_{ce}\Delta t/2)$
\citep{parker1991numerical,ricketson2020energy}, it can become much
smaller than the physical $\omega_{ce}$ for large $\Delta t$, effectively
eliminating the numerical stiffness from Bernstein waves \citep{stix1992waves}.
The resulting algorithm overcomes the particle-integration bottleneck
of the standard nonlinear implicit PIC for strongly magnetized plasmas
without needing to resolve the gyro-frequency.

The rest of the paper is organized as follows. Section \ref{sec:Methods}
introduces the method employed in this study. The numerical implementation
is described in Sec. \ref{sec:Numerical-implementation}. Section
\ref{sec:results} demonstrates the effectiveness of the proposed
algorithm with several electrostatic PIC tests. Finally, we conclude
in Sec. \ref{sec:conclusions}.

\section{Methodology}

\label{sec:Methods}We consider next the key components of our AP
particle-in-cell algorithm including particle substepping and orbit
averaging, treatment of reflective boundaries, field solve, and enforcement
of strict energy and charge conservation properties.

\textcolor{black}{To begin, we intend to solve the electrostatic Vlasov-Ampere
(VA) system for the electrostatic potential $\phi$ given by:
\begin{eqnarray}
\epsilon_{0}\frac{\partial\nabla^{2}\phi}{\partial t} & = & \nabla\cdot\mathbf{j},\label{eq:phi-eq}\\
\mathbf{j}(\mathbf{x},t) & = & \sum_{\alpha}q_{\alpha}\int d\mathbf{v}\mathbf{v}f_{\alpha}(\mathbf{x},\mathbf{v},t),\label{eq:current-eq}\\
\partial_{t}f_{\alpha}+\mathbf{v}\cdot\nabla f_{\alpha}+\frac{q_{\alpha}}{m_{\alpha}}\left(-\nabla\phi+\mathbf{v}\times\mathbf{B}_{0}\right)\cdot\nabla_{v}f_{\alpha} & = & 0.\label{eq:Vlasov-eq}
\end{eqnarray}
Assuming the usual particle ansatz for the species particle distribution
function, $f_{\alpha}\approx\sum_{p\in\alpha}w_{p}\delta(\mathbf{x}-\mathbf{x}_{p}(t))\delta(\mathbf{v}-\mathbf{v}_{p}(t))$,
gives the particle closure for the current density:
\begin{eqnarray}
\mathbf{j}(\mathbf{x},t) & = & \sum_{p}q_{p}w_{p}\mathbf{v}_{p}(t)\delta(\mathbf{x}-\mathbf{x}_{p}(t)),\label{eq:current-gather}\\
\frac{dw_{p}}{dt}=0 & ; & \frac{d\mathbf{x}_{p}}{dt}=\mathbf{v}_{p}\,\,\,\,;\,\,\,\,\frac{d\mathbf{v}_{p}}{dt}=\frac{q_{p}}{m_{p}}\left(-\nabla\phi+\mathbf{v}\times\mathbf{B}_{0}\right).\label{eq:pic-eom}
\end{eqnarray}
In this study, $\mathbf{B}_{0}$ is considered uniform. These equations
can be discretized with a multirate implicit formulation that conserves
local charge and total energy exactly, as described in Refs. \citep{chen2011energy,chen2014energy,chen2015multi}.
However, by design, the multirate particle integrator in the references
must resolve the local gyrofrequency for accuracy, incurring significant
inefficiencies for strongly magnetized plasmas. The purpose of this
study is to generalize these implicit algorithms to be able to use
arbitrary timesteps with respect to the gyrofrequency. In the very
large gyrofrequency limit, $\omega_{ce}\Delta t\gg1$, the particle
motion is described by its gyrocenter, and the current density moment
becomes the gyrocenter current contribution plus a magnetization current
\citep{brizard2016variational,hazeltine2018framework}:
\begin{equation}
\mathbf{j}=\mathbf{j}_{gc}+\nabla\times\mathbf{M},\label{eq:j_gPlusj_M}
\end{equation}
where $\mathbf{j}_{gc}$ is the gyro-center current. Since the magnetization
current is solenoidal, it follows that $\nabla\cdot\mathbf{j}=\nabla\cdot\mathbf{j}_{gc}$,
and therefore the field equation (Eq. \ref{eq:phi-eq}) remains invariant
in the strongly magnetized regime. This property renders the electrostatic
PIC system, Eqs. \ref{eq:phi-eq}, \ref{eq:current-gather}, and \ref{eq:pic-eom},
ideally suited for exploration of asymptotic preserving discretizations
of the particle-field equations.}

\subsection{Particle sub-stepping and orbit-averaging}

\label{subsec:pcle-substepping-orbavg}

We seek to employ very large timesteps compared to the gyroperiod
for advancing particles using a CN integrator (see Refs. \citep{chen2011energy,chen2014energy,chen2015multi}
and also the discussion below). By doing so, the particle orbits do
not resolve the gyromotion, but are still able to capture low-order
drift motions while preserving the Larmor radius \citep{ricketson2020energy}.
Specifically, as shown in the reference, CN without modification can
capture all the low-order drifts such as $E\times B$ drift, polarization
drift, curvature drift, etc., except for the magnetic drift (which
does not appear for a constant magnetic field, the case considered
in this study).

In earlier energy-conserving implicit PIC implementations \citep{chen2011energy,chen2014energy,chen2015multi},
automatic charge conservation was enforced without loss of energy
conservation by having particles stop at cell faces. While this approach
affords significant accuracy improvements vs. non-conserving strategies,
its disadvantage is that the number of cell crossings (and therefore
of substeps) increases with the size of the implicit timestep. The
cost of the orbit integration grows accordingly, thus offsetting any
potential efficiency gains of the implicit scheme from large timesteps.

To remove this limitation and improve efficiency without sacrificing
accuracy, we introduce in this study a new CN mover without the requirement
that each particle stops at cell faces, while still enforcing discrete
charge and energy conservation. The key innovation of the approach
is to allow orbit substeps that can be much larger than the cell size,
\emph{assuming each substep is straight with a constant velocity across
the whole} \emph{substep}. For an implicit field timestep $\Delta t$,
the orbit equations are solved for each orbit substep $\Delta\tau^{\nu}$
($\Delta t\geq\Delta\tau^{\nu}>0$, where $\nu$ denotes the substep
number), which is now determined according to physical considerations
beyond the cell size (e.g., see Ref. \citep{ricketson2020energy}).
Charge and energy conservation are still enforced strictly by keeping
track of segments (defined as the part of a substep that is within
a cell) crossed by the straight substep, a procedure done on the fly
in a very efficient manner. Accordingly, the average cost of each
particle orbit push becomes largely independent of the size of the
implicit field timestep, translating the efficiency savings directly
into wall-clock-time savings. In what follows, we outline the basic
particle sub-stepping and orbit-averaging techniques proposed in this
study.

\begin{figure}
\begin{centering}
\includegraphics{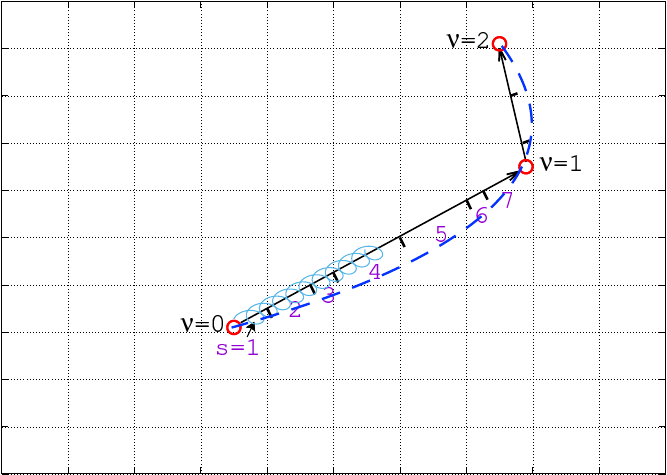}
\par\end{centering}
\caption{An illustration of a two-substep particle move. The substeps are between
for $\nu=0,1$ and $\nu=1,2$, each of which is a straight line. Within
each substep, the orbit line is further divided into segments separated
by cell boundaries, which are labeled by $s=1,2$, etc. as shown for
the first substep. The division into cell segments is done for each
substep. In our algorithm, depending on local conditions, it is possible
that a given substep approximates either an asymptotic guiding-center
orbit (e.g., the solid spiral orbit) or a part of a well-resolved
orbit (e.g., the dashed curve). \label{fig:substepping}}
\end{figure}

We begin with the implicit CN discretization for the particle orbit
equations used for each substep $\nu$,
\begin{align}
\mathbf{x}_{p}^{\nu+1} & =\mathbf{x}_{p}^{\nu}+\Delta\tau_{p}^{\nu}\mathbf{v}_{p}^{\nu+1/2},\label{eq:cn_x^nu}\\
\mathbf{v}_{p}^{\nu+1} & =\mathbf{v}_{p}^{\nu}+\frac{\Delta\tau_{p}^{\nu}q_{p}}{m_{p}}\left(\mathbf{E}_{p}^{\nu+1/2}+\mathbf{v}_{p}^{\nu+1/2}\times\mathbf{B}_{0}\right),\label{eq:cn_v^nu}
\end{align}
where $q_{p}$, $m_{p}$ are the particle charge and mass, the integer
superscript $\nu$ denotes the substep time level, $\mathbf{v}_{p}^{\nu+1/2}=(\mathbf{v}_{p}^{\nu}+\mathbf{v}_{p}^{\nu+1})/2$,
$\mathbf{B}_{0}$ is a constant external magnetic field (temporally
and spatially), $\mathbf{E}_{p}^{\nu+1/2}$ is the segment-averaged
particle electric field (to be specified), and $\Delta\tau_{p}^{\nu}$
is the particle sub-timestep. Note that, for any particle $p$, $\sum_{\nu\in n}\Delta\tau_{p}^{\nu}=\Delta t$,
where the summation is over all substeps in the timestep $n$. Equation
\ref{eq:cn_v^nu} can be analytically inverted \citep{buneman1980principles}
to give: 
\begin{align}
\mathbf{a}_{p} & =\mathbf{v}_{p}^{\nu}+\alpha_{p}\mathbf{E}_{p}^{\nu+1/2},\label{cn_inv1}\\
\mathbf{v}_{p}^{\nu+1/2} & ={\color{red}\frac{\mathbf{a}_{p}+\alpha_{p}\mathbf{a}_{p}\times\mathbf{B}_{0}+\alpha_{p}^{2}(\mathbf{a}_{p}\cdot\mathbf{B}_{0})\mathbf{B}_{0}}{1+(\alpha_{p}B_{0})^{2}},}\label{eq:cn_inv2}\\
\mathbf{x}_{p}^{\nu+1} & =\mathbf{x}_{p}^{\nu}+\Delta\tau_{p}^{\nu}\mathbf{v}_{p}^{\nu+1/2},\label{eq:cn_x_update_nu}\\
\mathbf{v}_{p}^{\nu+1} & =2\mathbf{v}_{p}^{\nu+1/2}-\mathbf{v}_{p}^{\nu},\label{cn_inv4}
\end{align}
where $\alpha_{p}=0.5\Delta\tau_{p}^{\nu}q_{p}/m_{p}$. Here, we have
assumed that, during the (possibly very large compared to the gyroperiod)
sub-timestep $\Delta\tau_{p}^{\nu}$, the velocity is constant and
the trajectory is a straight line (see Eq. \ref{eq:cn_x_update_nu}
and Fig. \ref{fig:substepping}). Since the right-hand-side of Eqs.
\ref{eq:cn_inv2}-\ref{eq:cn_x_update_nu} depends on both $\mathbf{x}_{p}^{\nu+1}$
and $\mathbf{v}_{p}^{\nu+1/2}$, a nonlinear iteration is needed.
However, the analytically inverted form in Eqs. \ref{cn_inv1}-\ref{cn_inv4}
is asymptotically well posed for large timesteps ($\alpha_{p}B_{0}=\Delta\tau_{p}\omega_{c,p}\gg1$),
and a simple Picard iteration generally converges to very tight tolerances
in very few iterations \citep{chen2014energy,koshkarov-jcp-2022-ap_pc}.

We now specify the particle electric field $\mathbf{E}_{p}^{\nu+1/2}$,
which is obtained by segment-averaging along the straight-line orbit
during the particle substep $\nu$ as: 
\begin{equation}
\mathbf{E}_{p}^{\nu+\nicefrac{1}{2}}=\sum_{h}\mathbf{E}_{h}^{n+\nicefrac{1}{2}}\cdot\left\langle \bar{\bar{\mathbf{S}}}(\mathbf{x}_{h}-\mathbf{x}_{p}^{s+\nicefrac{1}{2}})\right\rangle _{p}^{\nu},\label{eq:E-avg}
\end{equation}
where $h=(i,j,k)$ is the mesh index,
\begin{equation}
\mathbf{E}_{h}^{n+\nicefrac{1}{2}}=\frac{\mathbf{E}_{h}^{n+1}+\mathbf{E}_{h}^{n}}{2}\label{eq:E_h^1/2}
\end{equation}
is the mid-time electric field, defined at cell faces, the superscript
$s$ denotes a segment of length $\Delta\mathbf{x}_{p}^{s}$ on the
straight-line orbit segment $\Delta\mathbf{x}_{p}^{\nu}=\mathbf{x}_{p}^{\nu+1}-\mathbf{x}_{p}^{\nu}$
(see Fig. \ref{fig:substepping}), and $\mathbf{x}_{p}^{s+\nicefrac{1}{2}}$
is the mid-point of $\Delta\mathbf{x}_{p}^{s}$. In Eq. \ref{eq:E-avg},
we have introduced the segment-average operator of the shape function
$\left\langle \bar{\bar{\mathbf{S}}}(\mathbf{x}_{h}-\mathbf{x}_{p}^{s+\nicefrac{1}{2}})\right\rangle _{p}^{\nu}$
along the $\nu$-substep of the orbit of particle $p$, which will
be defined in detail below. In each segment $[\mathbf{x}_{p}^{s},\mathbf{x}_{p}^{s+1}]$,
the shape function dyad at cell faces is evaluated at $\mathbf{x}_{p}^{s+\nicefrac{1}{2}}$
as: 
\begin{eqnarray}
\bar{\bar{\mathbf{S}}}(\mathbf{x}_{i,j,k}-\mathbf{x}_{p}^{s+\nicefrac{1}{2}}) & = & \mathbf{i}\otimes\mathbf{i}\,\,S_{1}(x_{i+\nicefrac{1}{2}}-x_{p}^{s+\nicefrac{1}{2}})\mathcal{\mathbb{S}}_{22,jk}^{s+\nicefrac{1}{2}}(y_{p},z_{p})\nonumber \\
 & + & \mathbf{j}\otimes\mathbf{j}\,\,S_{1}(y_{j+\nicefrac{1}{2}}-y_{p}^{s+\nicefrac{1}{2}})\mathcal{\mathbb{S}}_{22,ik}^{s+\nicefrac{1}{2}}(z_{p},x_{p})\label{eq:S-doublebar}\\
 & + & \mathbf{k}\otimes\mathbf{k}\,\,S_{1}(z_{k+\nicefrac{1}{2}}-z_{p}^{s+\nicefrac{1}{2}})\mathcal{\mathbb{S}}_{22,ij}^{s+\nicefrac{1}{2}}(x_{p},y_{p}),\nonumber 
\end{eqnarray}
where $\mathbf{i},\mathbf{j},\mathbf{k}$ are unit vectors in the
$x$, $y$, $z$ directions, respectively, $\otimes$ denotes tensor
product, $S_{1}(x_{i+\nicefrac{1}{2}}-x_{p}^{s+\nicefrac{1}{2}})$
denotes linear B-spline shape function with $x_{i+\nicefrac{1}{2}}$
at cell faces, and
\begin{align}
\mathcal{\mathbb{S}}_{22,jk}^{s+\nicefrac{1}{2}}(y,z)\equiv\frac{1}{3} & \left[S_{2}(y_{j}-y^{s+1})S_{2}(z_{k}-z^{s+1})+\frac{S_{2}(y_{j}-y^{s})S_{2}(z_{k}-z^{s+1})}{2}\right.\nonumber \\
 & \left.\:\;+\frac{S_{2}(y_{j}-y^{s+1})S_{2}(z_{k}-z^{s})}{2}+S_{2}(y_{j}-y^{s})S_{2}(z_{k}-z^{s})\right]\label{eq:S_22}
\end{align}
is defined with the explicit purpose of enforcing charge conservation
\citep{stanier2019fully}, as will be shown below. In the last equation,
$S_{2}(y_{j}-y^{s+1})$ is the quadratic B-spline shape function,
with $y_{j}$ denoting a cell center (in $y$). The components of
the segment-averaged shape function dyad $\left\langle \bar{\bar{\mathbf{S}}}(\mathbf{x}_{h}-\mathbf{x}_{p}^{s+\nicefrac{1}{2}})\right\rangle _{p}^{\nu}$
along the $\nu$ substep of the orbit of particle $p$ are defined
as:
\begin{equation}
\mathbf{i}\cdot\left\langle \bar{\bar{\mathbf{S}}}(\mathbf{x}_{i,j,k}-\mathbf{x}_{p}^{s+\nicefrac{1}{2}})\right\rangle _{p}^{\nu}=\frac{\mathbf{i}}{\Delta x_{p}^{\nu}}\sum_{s\in\nu}S_{1}(x_{i+\nicefrac{1}{2}}-x_{p}^{s+\nicefrac{1}{2}})\mathcal{\mathbb{S}}_{22,jk}^{s+\nicefrac{1}{2}}(y_{p},z_{p})\Delta x_{p}^{s},\label{eq:orbit-average-nu}
\end{equation}
and similarly with $\mathbf{j}$ and $\mathbf{k}$ components. Here,
the summation is over all segments that belong to the substep $\nu$.

As a result of the above gathering, $\mathbf{E}_{p}^{\nu+\nicefrac{1}{2}}$
is the segment-averaged electric field for the straight line $[\mathbf{x}_{p}^{\nu},\mathbf{x}_{p}^{\nu+1}]$
of substep $\nu$. The number of $\Delta\mathbf{x}_{p}^{s}$ segments
is determined by the number of cell-crossings. The sub-timestep $\Delta\tau^{\nu}$
may be determined by a timestep estimator \citep{chen2011energy,ricketson2020energy},
as discussed below. Similarly, by symmetry (required for strict energy
conservation, as we show below), we obtain the segment-averaged current
density at cell faces from:
\begin{equation}
\mathbf{j}_{h,p}^{\nu+\nicefrac{1}{2}}=\frac{q_{p}}{\Delta_{h}}\mathbf{v}_{p}^{\nu+\nicefrac{1}{2}}\cdot\left\langle \bar{\bar{\mathbf{S}}}(\mathbf{x}_{h}-\mathbf{x}_{p}^{s+\nicefrac{1}{2}})\right\rangle _{p}^{\nu},\label{eq:current-density-nu}
\end{equation}
where $\Delta_{h}$ denotes cell volume.\textcolor{teal}{{} }\textcolor{red}{It
is worth noting that the proposed deposition scheme is similar to
the one proposed by Esirkepov in Ref. \citep{esirkepov2001exact},
but we allow segment averaging over many cells (instead of just one
in the reference), and we are limited to at most second-order B-splines
(instead of arbitrary order).}

\subsection{Boundary treatment of reflective particles}

\label{subsec:Boundary-treatment}

Realistic finite-domain particle simulations will feature \textcolor{red}{physical
}boundaries (e.g., walls), and therefore it is important to provide
a mechanism to adapt the AP particle orbit integration in their presence.
We consider here the case of perfectly reflecting walls, which do
not spoil energy conservation \citep{chacon2016curvilinear} and are
suitable to diagnose and demonstrate conservation properties. The
treatment proposed here can be generalized to other boundary conditions
(e.g., absorption, emission, etc.), if needed.

Given the particle sub-stepping strategy described above, it is natural
and straightforward to include a boundary condition. This is because
we can simply make one of the particle substeps end at the wall. This
is done by iterating the particle EOM such that the particle ends
at the wall when hitting it. For reflective boundary conditions, the
normal velocity of the particle is flipped when the particle hits
the wall. More specifically, if a particle is found to move beyond
a wall, we resolve the full particle orbit with the substep computed
as::
\begin{equation}
\Delta\tau_{p}^{\nu}=\textsf{min}\left(\Delta\tau_{p}^{\nu,*},\frac{x_{w}-x_{p}^{\nu}}{v_{px}^{\nu+\nicefrac{1}{2}}},\frac{y_{w}-y_{p}^{\nu}}{v_{py}^{\nu+\nicefrac{1}{2}}},\frac{z_{w}-z_{p}^{\nu}}{{\color{red}v_{pz}^{\nu+\nicefrac{1}{2}}}}\right),\label{eq:dtau_reflection}
\end{equation}
where $\Delta\tau_{p}^{\nu,*}$ is found from a timestep estimator
(see Ref. \citep{chen2011energy} and also below). If any velocity
component is zero, the particle will not move in that direction. Iterations
of the particle EOM for that substep will carry on with the updated
$\Delta\tau_{p}^{\nu}$ until convergence. In practice, we iterate
Eqs. \ref{cn_inv1}-\ref{eq:cn_x_update_nu} to convergence, and if
a particle intersects the boundary, its position for the end of that
substep is placed at the wall according to Eq. \ref{eq:dtau_reflection}.
After convergence, we flip the sign of its normal component to the
wall,
\[
\mathbf{n}\cdot\mathbf{v}_{p}^{\nu+1}\rightarrow-\mathbf{n}\cdot\mathbf{v}_{p}^{\nu+1},
\]
where $\mathbf{n}$ denotes the normal vector to the wall boundary,
and the orbit integration continues as before.

When the particle is deemed close to the boundary, its substep $\Delta\tau^{\nu}$
is estimated to control the integration error of the Crank--Nicolson
temporal scheme, which can be succinctly written as:
\begin{align}
\hat{\mathbf{x}}^{\nu+1} & =\mathbf{x}^{\nu}+\Delta\tau^{\nu}\frac{\hat{\mathbf{v}}^{\nu+1}+\mathbf{v}^{\nu}}{2},\label{eq:cn_x}\\
\hat{\mathbf{v}}^{\nu+1} & =\mathbf{v}^{\nu}+\Delta\tau^{\nu}\frac{\hat{\mathbf{a}}^{\nu+1}+\mathbf{a}^{\nu}}{2}.\label{eq:cn_v}
\end{align}
Assuming that $\mathbf{a}(\tau)\approx\mathbf{a}^{\nu}+(d\mathbf{a}/d\tau)^{\nu}\tau$,
we can re-write the equations as
\begin{equation}
\hat{\mathbf{x}}^{\nu+1}=\mathbf{x}^{\nu}+\mathbf{v}^{\nu}\Delta\tau^{\nu}+\frac{1}{2}\mathbf{a}^{\nu}(\Delta\tau^{\nu})^{2}+\frac{1}{4}(\frac{d\mathbf{a}}{d\tau})^{\nu}(\Delta\tau^{\nu})^{3}+\ensuremath{\mathcal{O}}((\Delta\tau^{\nu})^{4}).\label{eq:x_cn}
\end{equation}
Taylor-expanding the exact solution in time we obtain:
\begin{equation}
\mathbf{x}^{\nu+1}=\mathbf{x}^{\nu}+\mathbf{v}^{\nu}\Delta\tau^{\nu}+\frac{1}{2}\mathbf{a}^{\nu}(\Delta\tau^{\nu})^{2}+\frac{1}{6}(\frac{d\mathbf{a}}{d\tau})^{\nu}(\Delta\tau^{\nu})^{3}+\ensuremath{\mathcal{O}}((\Delta\tau^{\nu})^{4}).\label{eq:x_Taylor}
\end{equation}
The leading local truncation error is found by subtracting Eq. \ref{eq:x_Taylor}
and \ref{eq:x_cn}, yielding:
\begin{equation}
\mathcal{E}_{\Delta\tau}=\frac{1}{12}(\frac{d\mathbf{a}}{d\tau})^{\nu}(\Delta\tau^{\nu})^{3}.\label{eq:local_error_cn}
\end{equation}
Equation \ref{eq:local_error_cn} is rearranged to find an estimate
of $\Delta\tau^{\nu}$ as:
\begin{equation}
\Delta\tau^{\nu}=\left(\frac{12\mathcal{E}_{\Delta\tau}}{\left\Vert (d\mathbf{a}/d\tau)^{\nu}\right\Vert }\right)^{\frac{1}{3}},\label{eq:delta-tau-estimate}
\end{equation}
where $\mathcal{E}_{\Delta\tau}$ is a user-provided local error tolerance
(here we use $\mathcal{E}_{\Delta\tau}=0.01$). The rate of change
of the acceleration needs to be estimated along the particle substep
$\nu$. In this study, we predict it at the beginning of the substep
by using Euler's scheme with a tiny step $\delta\tau=10^{-8}$ to
push the particle to a new location $\mathbf{x}_{E}$, and compute
$(\frac{d\mathbf{a}}{d\tau})^{\nu}\simeq(\mathbf{a}(\mathbf{x}_{E})-\mathbf{a}^{\nu})/\delta\tau$.
More sophisticated estimates that account for the change of $\mathbf{a}$
along the full substep may be needed in the future, and will be explored
in future work.

\subsection{Field solver\label{subsec:Field-solver}}

We find the electrostatic potential from Gauss' law combined with
the continuity equation \citep{chen2014energy,chen2015multi}:
\begin{equation}
\epsilon_{0}\frac{\partial\nabla^{2}\phi}{\partial t}-\nabla\cdot\mathbf{j}=0.\label{eq:Ampere-Poisson}
\end{equation}
In discrete form, it reads:
\begin{equation}
\epsilon_{0}(\nabla_{h}^{2}\phi_{h}^{n+1}-\nabla_{h}^{2}\phi_{h}^{n})-\Delta t\nabla_{h}\cdot\bar{\mathbf{j}}_{h}^{n+1/2}=0,\label{eq:Ampere-Poisson-d}
\end{equation}
which is the standard second-order time-domain finite-difference scheme
on the Yee mesh (as in Ref. \citep{chen2015multi}), with $\phi_{h}$
defined at cell centers and components of $\bar{\mathbf{j}}_{h}$
colocated with the electric field ones (normal to faces). The orbit-averaged
current density is obtained from:
\begin{equation}
\bar{\mathbf{j}}_{h}^{n+1/2}=\frac{1}{\Delta t}\sum_{p}\sum_{\nu\in n}\mathbf{j}_{h,p}^{\nu+\nicefrac{1}{2}}\Delta\tau_{p}^{\nu},\label{eq:orbit-average-j}
\end{equation}
which can be expanded using Eq. \ref{eq:current-density-nu} as:
\begin{align}
\bar{\mathbf{j}}_{h}^{n+1/2} & =\frac{1}{\Delta t}\sum_{p}\sum_{\nu\in n}\frac{q_{p}}{\Delta_{h}}\mathbf{v}_{p}^{\nu+\nicefrac{1}{2}}\cdot\left\langle \bar{\bar{\mathbf{S}}}(\mathbf{x}_{h}-\mathbf{x}_{p}^{s+\nicefrac{1}{2}})\right\rangle ^{\nu}\Delta\tau_{p}^{\nu},\nonumber \\
 & =\frac{1}{\Delta_{h}\Delta t}\sum_{p}\sum_{s\in n}q_{p}\Delta\mathbf{x}_{p}^{s}\cdot\bar{\bar{\mathbf{S}}}(\mathbf{x}_{h}-\mathbf{x}_{p}^{s+\nicefrac{1}{2}}),\label{eq:orbit-average-j-expanded}
\end{align}
where the double sum performs the average over all segments of all
particles for the whole timestep. The electric field is obtained at
integer time levels and cell faces by a central-difference face gradient:
\begin{equation}
\mathbf{E}_{h}=-\nabla_{h}\phi_{h}.\label{eq:find_Eh}
\end{equation}
As is demonstrated in the next section, the discrete continuity equation
\begin{equation}
\rho_{h}^{n+1}-\rho_{h}^{n}+\Delta t\nabla_{h}\cdot\bar{\mathbf{j}}_{h}^{n+1/2}=0\label{eq:continuity-d}
\end{equation}
 is exactly satisfied when $\rho_{h}$ is defined as:
\begin{equation}
\rho_{h}=\sum_{p}\frac{q_{p}}{\Delta_{h}}S_{2}(x_{i+1/2}-x_{p})S_{2}(y_{j+1/2}-y_{p})S_{2}(z_{k+1/2}-z_{p}).\label{eq:rho-d}
\end{equation}
Therefore, up to the nonlinear tolerance, the discrete Gauss's law
follows from Eq. \ref{eq:Ampere-Poisson-d} and \ref{eq:continuity-d}
as
\begin{equation}
\epsilon_{0}\nabla_{h}^{2}\phi_{h}^{n+1}=-\rho_{h}^{n+1},\label{eq:Ampere-Poisson-d-1}
\end{equation}
for $n>0$ if it is satisfied at $n=0$.

\subsection{Energy and charge conservation properties\label{subsec:Energy-and-charge}}

We show that the proposed algorithm is energy- and charge-conserving
as follows. For energy conservation, we prove that the discrete total
energy (TE) is exactly conserved without energy sources, i.e.,
\begin{equation}
\left.\left(\mathrm{TE}\right)\right|_{n}^{n+1}=\left.\left(W_{E}+K\right)\right|_{n}^{n+1}=0,\label{eq:TE}
\end{equation}
where
\[
W_{E}^{n}\equiv\frac{\epsilon_{0}}{2}\sum_{h}\bm{\Delta}_{h}(\mathbf{E}_{h}^{n})^{2},
\]
and 
\[
K^{n}\equiv\frac{1}{2}\sum_{p}m_{p}(v_{p}^{n})^{2}.
\]
This is shown as follows:
\begin{align*}
K^{n+1}-K^{n} & =\sum_{p}m_{p}\frac{\mathbf{v}_{p}^{n+1}+\mathbf{v}_{p}^{n}}{2}(\mathbf{v}_{p}^{n+1}-\mathbf{v}_{p}^{n})\\
 & =\sum_{p}\cancel{m_{p}}\sum_{\nu\in n}\frac{\mathbf{v}_{p}^{\nu+1}+\mathbf{v}_{p}^{\nu}}{2}\cdot\left(\mathbf{E}_{p}^{\nu+1/2}+\frac{\mathbf{v}_{p}^{\nu+1}+\mathbf{v}_{p}^{\nu}}{2}\times\mathbf{B}_{0}\right)\frac{\Delta\tau^{\nu}q_{p}}{\cancel{m_{p}}}\\
 & =\sum_{p}\sum_{\nu\in n}\Delta\mathbf{x}_{p}^{\nu}q_{p}\sum_{h}\mathbf{E}_{h}^{n+\nicefrac{1}{2}}\cdot\left\langle \bar{\bar{\mathbf{S}}}(\mathbf{x}_{h}-\mathbf{x}_{p}^{s+\nicefrac{1}{2}})\right\rangle _{p}^{\nu}\\
 & =\sum_{h}\mathbf{E}_{h}^{n+\nicefrac{1}{2}}\cdot\bar{\mathbf{j}}_{h}^{n+1/2}\Delta_{h}\Delta t\\
 & =-\sum_{h}\nabla_{h}\phi_{h}^{n+\nicefrac{1}{2}}\cdot\bar{\mathbf{j}}_{h}^{n+1/2}\Delta_{h}\Delta t\\
 & =\sum_{h}\phi_{h}^{n+\nicefrac{1}{2}}\nabla_{h}\cdot\bar{\mathbf{j}}_{h}^{n+1/2}\Delta_{h}\Delta t\\
 & =\sum_{h}\phi_{h}^{n+\nicefrac{1}{2}}\epsilon_{0}\frac{\nabla_{h}^{2}\phi_{h}^{n+1}-\nabla_{h}^{2}\phi_{h}^{n}}{\cancel{\Delta t}}\Delta_{h}\cancel{\Delta t}\\
 & =-\frac{\epsilon_{0}}{2}\sum_{h}\left[(\mathbf{E}_{h}^{n+1})^{2}-(\mathbf{E}_{h}^{n})^{2}\right]\Delta_{h}\\
 & =-(W_{E}^{n+1}-W_{E}^{n}),
\end{align*}
where we have used Eq. \ref{eq:cn_v^nu} for the second equality,
Eq. \ref{eq:cn_x^nu}, \ref{eq:E-avg} and $\frac{\mathbf{v}^{\nu+1}+\mathbf{v}^{\nu}}{2}\cdot\left(\frac{\mathbf{v}^{\nu+1}+\mathbf{v}^{\nu}}{2}\times\mathbf{B}_{0}\right)=0$
for the third equality, Eq. \ref{eq:current-density-nu}, \ref{eq:orbit-average-j}
for the fourth equality, Eq. \ref{eq:find_Eh} for fifth equality,
integration by parts in the sixth equality, Eq. \ref{eq:Ampere-Poisson-d}
for the seventh equality, integration by parts and Eq. \ref{eq:find_Eh}
for the eighth equality, and we denote $\phi^{n+\nicefrac{1}{2}}\equiv(\phi^{n+1}+\phi^{n})/2$.
Note that the discrete integration (or summation) by parts is satisfied
for the standard second-order finite-difference schemes with periodic
boundary conditions. We note that the derivation remains valid with
a perfect-conductor boundary and reflecting particles. Assuming that
we have perfect conductor walls confining the $y$-direction, with
$\phi=0$ at both walls (i.e., no imposed electric field), all the
above steps remain the same, including the summation by parts:
\[
\sum_{j=0}^{N_{y}+1}\frac{\phi_{j}-\phi_{j-1}}{\Delta y}\bar{\mathfrak{j}}_{j}=-\sum_{j=0}^{N_{y}+1}\phi_{j}\frac{\bar{\mathfrak{j}}_{j}-\bar{\mathfrak{j}}_{j-1}}{\Delta y},
\]
where we have used $\phi_{0}=\phi_{N_{y+1}}=0$. In multiple dimensions,
the summation by parts is done for each dimension separately.

Regarding charge conservation, we show next that
\begin{equation}
\rho_{h}^{n+1}-\rho_{h}^{n}+\Delta t\nabla_{h}\cdot\mathbf{\bar{j}}_{h}^{n+1/2}=0.\label{eq:disc-cc}
\end{equation}
It is sufficient that Eq. \ref{eq:disc-cc} be satisfied for each
particle (the total follows when summing over all particle contributions),
which can be written as
\begin{equation}
\sum_{\nu\in n}\left(\rho_{h,p}^{\nu+1}-\rho_{h,p}^{\nu}+\Delta\tau_{p}^{\nu}\nabla_{h}\cdot\mathbf{j}_{h,p}^{\nu+\nicefrac{1}{2}}\right)=0.\label{eq:disc-cc-nu}
\end{equation}
It is therefore sufficient for the continuity equation to be satisfied
for each substep $\nu$, which may be written as
\begin{equation}
\sum_{s\in\nu}(\rho_{h,p}^{s+1}-\rho_{h,p}^{s})+\Delta\tau_{p}^{\nu}\nabla_{h}\cdot\mathbf{j}_{h,p}^{\nu+\nicefrac{1}{2}}=0,\label{eq:disc-cc-nu-s}
\end{equation}
where:
\begin{equation}
\rho_{h,p}^{s}=\frac{q_{p}}{\Delta_{h}}S_{2}(x_{i}-x_{p}^{s})S_{2}(y_{j}-y_{p}^{s})S_{2}(z_{k}-z_{p}^{s}),\label{eq:rho-cc-nu-s}
\end{equation}
and as before we use $s$ and $s+1$ for the start and end points
of a segment $s$, respectively. Equation \ref{eq:disc-cc-nu-s} can
be expanded using Eq. \ref{eq:current-density-nu} as:
\begin{equation}
\sum_{s\in\nu}\left[\rho_{h,p}^{s+1}-\rho_{h,p}^{s}+\nabla_{h}\cdot\frac{q_{p}}{\Delta_{h}}\bar{\bar{\mathbf{S}}}(\mathbf{x}_{h}-\mathbf{x}_{p}^{s+\nicefrac{1}{2}})\cdot\Delta\mathbf{x}_{p}^{s}\right]=0.\label{eq:cc-nu-s}
\end{equation}
The expression in the square bracket describes the continuity equation
for each segment $s$. Given Eqs. \ref{eq:rho-cc-nu-s} and \ref{eq:S-doublebar},
\ref{eq:S_22}, it follows that the continuity equation is exactly
satisfied because:
\begin{eqnarray}
(\rho_{h,p}^{s+1}-\rho_{h,p}^{s})\Delta_{h}/q_{p} & = & S_{2}(x_{i}-x_{p}^{s+1})S_{2}(y_{j}-y_{p}^{s+1})S_{2}(z_{k}-z_{p}^{s+1})\nonumber \\
 & - & S_{2}(x_{i}-x_{p}^{s})S_{2}(y_{j}-y_{p}^{s})S_{2}(z_{k}-z_{p}^{s})\nonumber \\
 & = & \left[S_{2}(x_{i}-x_{p}^{s+1})-S_{2}(x_{i}-x_{p}^{s})\right]\mathcal{\mathbb{S}}_{22,jk}^{s+\nicefrac{1}{2}}(y_{p},z_{p})\nonumber \\
 & + & \left[S_{2}(y_{j}-y_{p}^{s+1})-S_{2}(y_{j}-y_{p}^{s})\right]\mathbb{S}_{22,ik}^{s+\nicefrac{1}{2}}(z_{p},x_{p})\nonumber \\
 & + & \left[S_{2}(z_{k}-z_{p}^{s+1})-S_{2}(z_{k}-z_{p}^{s})\right]\mathbb{S}_{22,ij}^{s+\nicefrac{1}{2}}(x_{p},y_{p}),\label{eq:drho}
\end{eqnarray}
and, in the $x$-direction, it can be shown that \citep{chen2011energy}:
\begin{equation}
S_{2}(x_{i}-x_{p}^{s+1})-S_{2}(x_{i}-x_{p}^{s})+(\Delta x_{p}^{s})\frac{S_{1}(x_{i+\nicefrac{1}{2}}-x_{p}^{s+\nicefrac{1}{2}})-S_{1}(x_{i-\nicefrac{1}{2}}-x_{p}^{s+\nicefrac{1}{2}})}{\Delta x}=0,\label{eq:cc-1d-1}
\end{equation}
and similarly for the $y$ and $z$ directions. Substituting Eq. \ref{eq:drho}
in Eq. \ref{eq:cc-nu-s} proves that Eq. \ref{eq:cc-nu-s} is exactly
satisfied, and therefore so is Eq. \ref{eq:disc-cc}. As with earlier
charge-conserving implementations \citep{chen2011energy,chen2015multi},
$\rho$ and $\mathbf{j}$ can use lower-order shape functions if desired
(see Ref. \citep{chen2020semi} for an example).

\section{Numerical implementation\label{sec:Numerical-implementation}}

We describe the full algorithm in detail in Algorithm \ref{alg:Iterative-solution-of-CN},
including the outer (field) time step (line 5), the residual evaluation
(line 15), and the particle push (line 20). We describe the particle
and field iterative algorithms in some detail next.
\begin{algorithm}
\caption{\label{alg:Iterative-solution-of-CN}Fully implicit time-advance algorithm
for the asymptotic-preserving particle-in-cell algorithm proposed
in this study.}

\begin{algorithmic}[1] 
\State{Given initial $\phi_h^0$ and $\left\{ \mathbf{x}_p^{n},\mathbf{v}_p^{n}\right\} $}
\For{$n=0$ to $N-1$}  \Comment{\textsf{$N$: total number of timesteps}}
\State{$\phi_h^{n+1} = \textsf{advancePotential} (\phi_h^{n})$}
\EndFor
\Statex
\Procedure{advancePotential}{$\phi_h^n$}
\State{$\phi_h^{n+1} \leftarrow  \phi_h^{n}$}
\State{$R_0 \leftarrow \textsf{evaluateResidual}(\phi_h^{n+1}); ~ R = R_0$} 
\While{$||R||>\epsilon_a + \epsilon_r ||R_0||$} \Comment{\textsf{$\epsilon_{a,r}\equiv$ nonlinear tolerances}} 
\State{$\delta \phi_h^{n+1}=-P^{-1}R$ \Comment{Apply preconditioner \cite{chen2015multi}}}
\State{$\delta \phi_h^{n+1} \leftarrow \textsf{AA}(\delta \phi_h^{n+1},\textsf{nvec})$ \Comment{Apply Anderson Acceleration}} 
\State{ $\phi_h^{n+1}\leftarrow\phi_h^{n+1}+\delta \phi_h^{n+1}$ \Comment{Update nonlinear state}} 
\State{$R \leftarrow \textsf{evaluateResidual}(\phi_h^{n+1})$} 
\EndWhile 
\EndProcedure
\Statex
\Procedure{evaluateResidual} {$\phi_h^{n+1}$}
\State{$\mathbf{E}_h^{n+\nicefrac{1}{2}}\leftarrow-\nabla_h(\phi_h^{n+1}+\phi_h^n)/2$}
\State{$\bar{\mathbf{j}}_h^{n+1/2} \leftarrow$  pushParticles ($\mathbf{E}_h^{n+\nicefrac{1}{2}}, \mathbf{B}_0) $ }

\State{$R = \epsilon_{0}(\nabla_{h}^{2}\phi_{h}^{n+1}-\nabla_{h}^{2}\phi_{h}^{n})-\Delta t\nabla_{h}\cdot\bar{\mathbf{j}}_{h}^{n+1/2}$ \Comment{Eq.~\ref{eq:Ampere-Poisson-d}}}
\EndProcedure
\Statex

\Procedure{pushParticles} {$\mathbf{E}_h^{n+\nicefrac{1}{2}}$ , $\mathbf{B}_0$}
\For{each particle} 
\State{$\tau\leftarrow 0$ ; $\nu\leftarrow 0$}
\State{$\mathbf{x}_p^\nu \leftarrow \mathbf{x}_p^n$ ; $\mathbf{v}_p^\nu \leftarrow \mathbf{v}_p^n$}
\While{$\tau <\Delta t$} 
\State{Estimate $\Delta \tau_p^\nu$}\Comment{{e.g., $\Delta\tau_p^\nu=\Delta t$ or Eq. \ref{eq:delta-tau-estimate}}}
\State{Set $\mathbf{x}_p^{\nu+1} \leftarrow \mathbf{x}_p^{\nu}$ ; $\mathbf{v}_p^{\nu+1} \leftarrow \mathbf{v}_p^{\nu}$ }
	\While{not converged} 
		\State{Compute segments $\Delta \mathbf{x}_s^\nu$ and perform field average \Comment{Eq.~\ref{eq:E-avg}}}
		\State{Solve for \{$\mathbf{x}_p^{\nu+1},\mathbf{v}_p^{\nu+1}$\} \Comment{Eqs.~\ref{cn_inv1}-\ref{cn_inv4}}}
	\EndWhile
\State{Recompute segments $\Delta \mathbf{x}_s^\nu$ and collect $\mathbf{j}_{h,p}^{\nu+\nicefrac{1}{2}}$ \Comment{Eq.~\ref{eq:current-density-nu}}}
\State{$\tau\leftarrow \tau+\Delta \tau_p^\nu$}
\State{$\nu\leftarrow \nu+1$}
\EndWhile
\State{Collect $\bar{\mathbf{j}}_h$ \Comment{Eq.~\ref{eq:orbit-average-j}}}
\EndFor
\EndProcedure
\end{algorithmic}
\end{algorithm}

\subsection{Particle iterative algorithm (procedure \noun{pushParticles} in Alg.
\ref{alg:Iterative-solution-of-CN})}

The procedure \noun{pushParticles} consists of a triple loop, the
outer of which cycles over particles, the middle one traverses the
orbit in substeps, and the inner one converges on the particle equations
per substep. In the inner loop, we solve the orbit equations for each
particle, which are iterated to convergence in each substep $\nu$,
with electric fields and current densities segment-averaged as indicated
earlier. Substepping is equipped with a timestep estimator for $\Delta\tau_{p}$
(Eq. \ref{eq:delta-tau-estimate}). Particle substepping can be particularly
useful in the presence of magnetic field gradients \citep{ricketson2020energy},
when the magnetic field is weak, and for boundary treatment of particles
near a wall. In the present setting, we initially set $\Delta\tau_{p}^{\nu}=\Delta t$
for all particles, and only turn on the timestep estimator when convergence
for a given orbit substep is slow or when near a physical boundary.

The particle phase-space positions per substep $\nu$ are found by
iterating Eqs. \ref{eq:cn_x^nu}-\ref{eq:cn_v^nu} in a Picard fashion.
The iteration is initialized with the previous substep position and
velocity, $(\mathbf{x}_{p}^{\nu},\mathbf{v}_{p}^{\nu})$. After each
substep iteration to find the final position $(\mathbf{x}_{p}^{\nu+1},\mathbf{x}_{p}^{\nu+1})$,
segments $\Delta\mathbf{x}_{p}^{s}$ are computed from cell crossings
assuming a straight orbit (Fig. \ref{fig:substepping}). Once segments
are known, the electric field is segment-averaged (Eq. \ref{eq:E-avg}),
and the iteration proceeds.

The segment-averaged current density is obtained once the substep
iteration is complete. The average particle electric field (Eq. \ref{eq:E-avg})
is gathered to particles first at each $\mathbf{x}_{p}^{s+1/2}$ location
(involving a sum over adjacent cells), and then averaged along the
corresponding $\nu$ substep. The current density scatter is implemented
exactly as indicated in Eq. \ref{eq:orbit-average-j-expanded}, i.e.,
accumulated first at each cell per segment per particle, and then
adding all particle's contributions in every cell. This procedure
ensures that all averaging operations are performed on a per particle
basis and locally on the mesh.

\subsection{Field iterative algorithm (procedure \noun{advancePotential} in Alg.
\ref{alg:Iterative-solution-of-CN})}

\label{subsec:Iterative-algorithm}

The procedure \noun{advancePotential} in Algorithm \ref{alg:Iterative-solution-of-CN}
requires an iterative nonlinear solver. Here, we employ a preconditioned
Anderson Acceleration (AA) solver \citep{anderson1965iterative,walker2011anderson}.
As opposed to Jacobian-free Newton-Krylov (JFNK) solvers, AA does
not require differentiation and is therefore less susceptible to convergence
issues due to non-differentiability in particle orbits (which occur,
for instance, when particle orbits diverge due to the perturbation
in the Gateaux derivative in JFNK). The nonlinear residual that drives
the AA is formulated by enslaving the particle-push step to obtain
the current density (as indicated in the \noun{evaluateResidual} procedure
in Alg. \ref{alg:Iterative-solution-of-CN}). We precondition the
iteration by using the electrostatic component of the more general
electromagnetic preconditioner discussed in Ref. \citep{chen2015multi}.

\section{Numerical experiments}

\label{sec:results}In what follows, we consider three different problems
on varying complexity and magnetization: a single-magnetized-species
Diocotron instability in 2D, a 1D modified two-stream instability
in which electrons are magnetized but ions are not, and a drift-wave
instability with both ions and electrons magnetized in 2D with perfect-conductor
boundaries. Unless otherwise stated, we normalize the plasma quantities
to ion units, in which the charge-mass ratio $q_{i}/m_{i}$, ion plasma
frequency $\omega_{pi}$, and ion Debye length $\lambda_{Di}$ are
set to unity.

\subsection{Diocotron instability}

\begin{figure}
\centering{}\includegraphics[width=1\columnwidth]{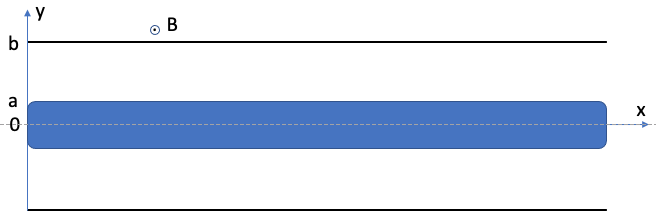}\caption{Simulation setup for the Diocotron instability. The 2D domain is periodic
in $x$ and confined in $y$. An electron beam of a width $a$ is
sitting in the middle of the domain. A uniform magnetic field B is
pointing in the $z$ direction, perpendicular to the $x-y$ plane.\label{fig:Diocotron-setup}}
\end{figure}

The Diocotron instability occurs in a low density ($\omega_{p}<\omega_{c}$)
cross-field electron population. Here, we follow the simulation setup
of Ref. \citep{levy1968computer}. A cold electron beam of uniform
density is placed in between two conducting plates without contact.
There are no ions in this simulation, so we use electron units: $q_{e}/m_{e}=\omega_{pe}=\lambda_{De}=1$.
The magnetic field is uniform in the $z$-direction, and of magnitude
such that $\omega_{ce}/\omega_{pe}=\sqrt{20}$ (i.e., $B_{z}=\sqrt{20}$).
The simulation domain is $[-8,8]\times[-8,8]\lambda_{De}^{2}$ in
2D, with periodic boundary conditions (B.C.) in the $x$-direction
and perfect conductor and reflective B.C. in the $y$-direction. The
electrons are stationary to begin with, and uniformly distributed
in $[-8,8]\times[-1,1]\lambda_{De}^{2}$ (see Fig. \ref{fig:Diocotron-setup}).
The simulation is performed in a 2D-3V configuration using a $32\times64$
grid and 40 particles per cell.

\begin{figure}
\centering{}\includegraphics[width=1\columnwidth]{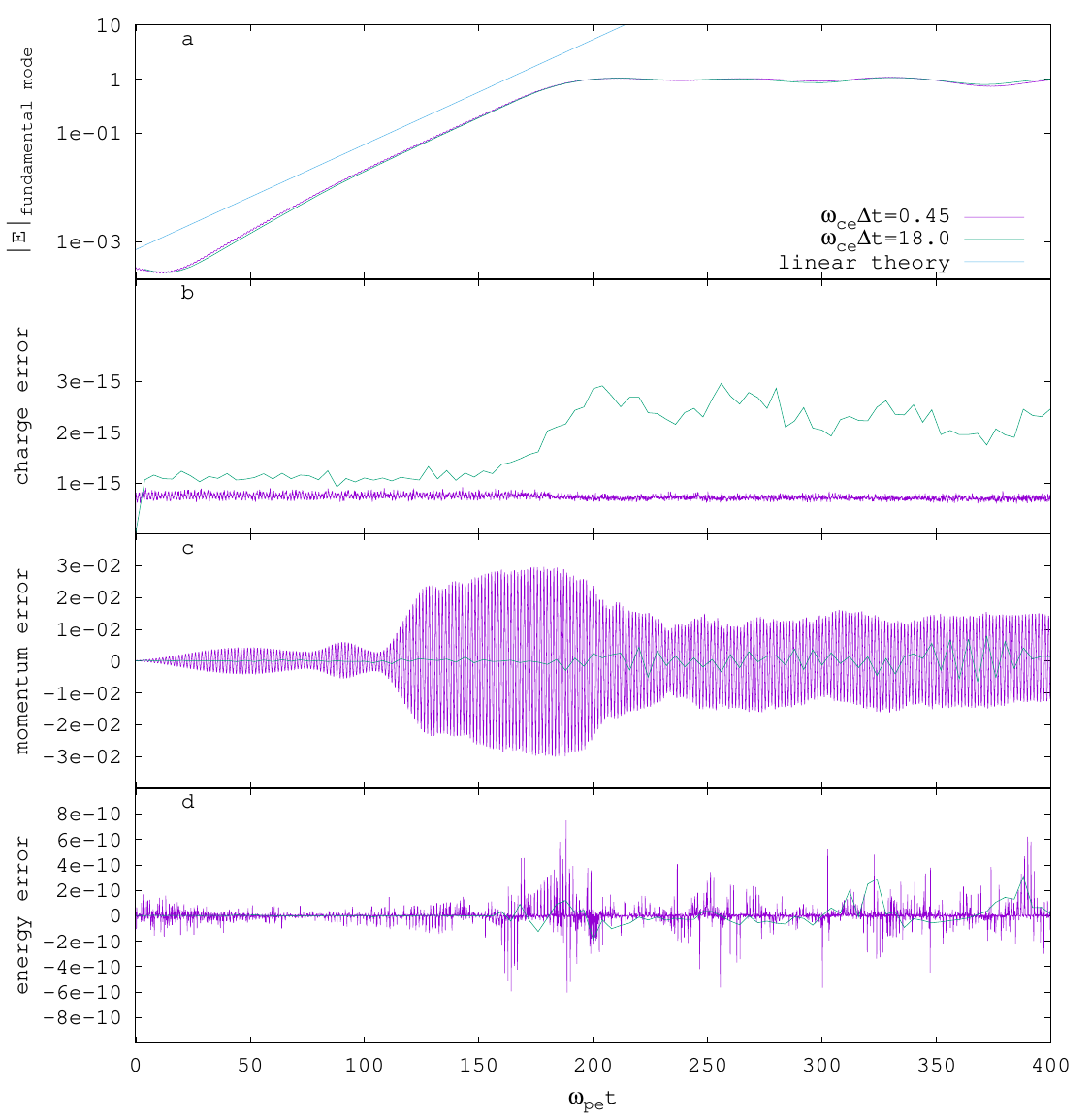}\caption{Simulation results for the Diocotron instability problem. Panel a
depicts the growth in amplitude of the fundamental mode ($k=2\pi/L_{x}$).
Panels b, c, d are time histories of the root-mean-square (rms) of
the charge continuity equation (defined as root-mean-square of the
residual of Eq. \ref{eq:disc-cc}) , the total momentum error (defined
as sum of all particles momentum in the $x$-direction) and total
energy error (defined as $(TE^{n+1}-TE^{n})/TE^{n}$), respectively.\label{fig:Diocotron}}
\end{figure}

Figure \ref{fig:Diocotron} shows simulation results for the Diocotron
instability using the CN pusher proposed here. The results in the
figure demonstrate excellent agreement in the linear growth rate between
theory and simulations for both a relatively small timestep ($\omega_{ce}\Delta t=0.45$,
which is comparable to that used in Ref. \citep{levy1968computer})
and a very large one ($\omega_{ce}\Delta t=18$). Agreement is also
excellent in the nonlinear stage of the simulation. Proof of charge
conservation (to numerical round off) and energy conservation (to
nonlinear tolerance) is also provided (panels b and d). Momentum is
not exactly conserved in either simulation, but errors are much smaller
with the AP integrator than the well resolved one (panel c). The growth
rate of the fundamental mode is $\gamma=0.2\omega_{0}=0.01\omega_{ce}=0.045\omega_{pe}$
\citep{levy1968computer}, where $\omega_{0}=d(E/B)/dy$ is the characteristic
timescale of the shear in the $\mathbf{E}\times\mathbf{B}$ drift
velocity. Therefore, the largest timestep considered is $\Delta t\gamma=0.18$,
which is still reasonable for a second-order method. The total number
of nonlinear iterations per time step increases only by a factor of
2 (from 5 to 10) when increasing $\Delta t$ by a factor of 40 for
a nonlinear tolerance of $10^{-10}$, resulting in an overall CPU
speedup of about 14.5.

\subsection{Modified two-stream instability}

\begin{figure}
\centering{}\includegraphics[width=1\columnwidth]{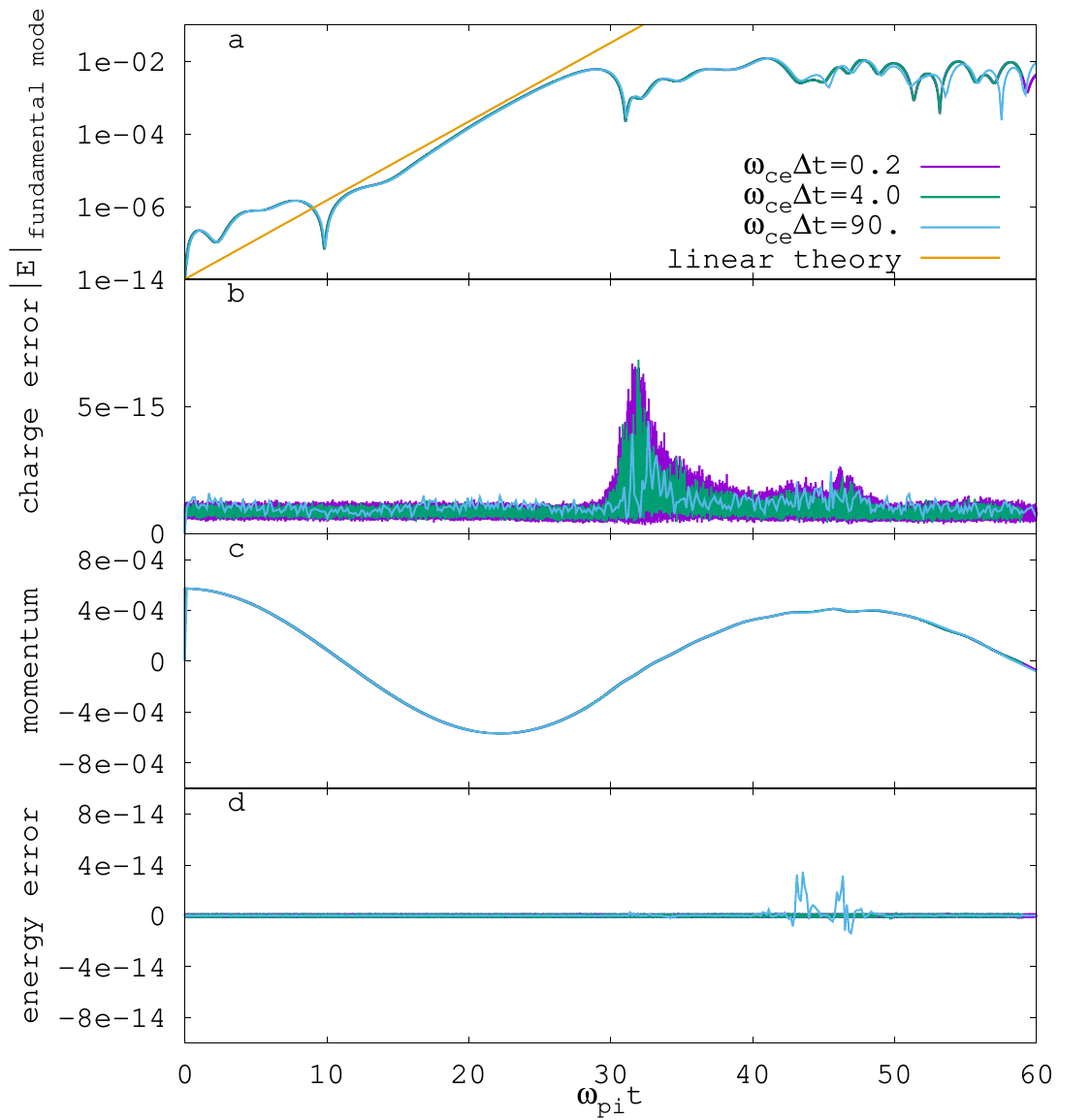}\caption{Simulation results for the modified two-stream instability problem.
Panel a depicts the growth in the electric-field energy with three
different timesteps. Panels b, c, d depict time histories of the root-mean-square
(rms) of the charge continuity equation, the total momentum (normalized
by $\sum_{p}m_{i}U$) , and total energy error, respectively (measured
as in Fig. \ref{fig:Diocotron}). Nonlinear tolerance is set at $10^{-10}$.\label{fig:modified-2stream}}
\end{figure}
Figure \ref{fig:modified-2stream} shows results from a simulation
of the modified two-stream instability (MTSI) \citep{mcbride1972theory},
in which electrons and ions are drifting towards each other across
an external magnetic field. The MTSI happens in the regime $\omega_{ce}\gg\omega\gg\omega_{ci}$,
i.e., electrons are strongly magnetized but ions are unmagnetized.
The cold plasma dispersion relation is \citep{mcbride1972theory}:
\[
1+\sin(\theta)\frac{\omega_{pe}^{2}}{\omega_{ce}^{2}}-\frac{1}{(\omega-\mathbf{k}\cdot\mathbf{U})^{2}}-\cos(\theta)\frac{\omega_{pe}^{2}}{\omega^{2}}=0,
\]
where $\theta$ is a small angle between the wave vector $\mathbf{k}$
and the relative velocity $\mathbf{U}$, and the magnetic field $\mathbf{B}$
is perpendicular to $\mathbf{U}$. The simulation is performed in
a 1D-3V configuration with $L_{x}=1.8229$ $\lambda_{Di}$, 32 cells
and 100 particles per cell. In this test, $\omega_{ce}/\omega_{pe}=10$
and $m_{i}/m_{e}=5000$. In ion units, $m_{i}=1.0$ and $B=1/\sqrt{50}$.
The magnetic field is mostly pointed in the $y$ direction but slightly
tilted such that $B_{x}/B=\sin(\theta)=\sqrt{m_{e}/m_{i}}$. Electrons
are stationary, and ions have a relative velocity of $U=0.5$. For
these parameters, the growth rate is $\gamma=0.4992\omega_{pi}.$

We consider three timestep sizes, a small one with about 32 steps
per gyroperiod ($\Delta t=0.2\omega_{ce}^{-1}$), an intermediate
one with about 1.6 steps per gyro-period ($\Delta t=4.0\omega_{ce}^{-1}$),
and a very large one with 0.07 steps per gyro-period ($\Delta t=90\omega_{ce}^{-1}$).
The results in the figure again demonstrate excellent agreement for
all timesteps, and excellent conservation of charge and energy (panels
b and d).

\begin{table}
\centering{}\caption{Solver performance for the modified two-stream instability problem
with respect to mass ratio $m_{i}/m_{e}$ for the largest timestep
considered. The problem is set up using ion units, so the only parameter
that is varied is the electron mass. All runs employ a fixed timestep
$\Delta t=0.127\omega_{pi}^{-1}$, corresponding to different timesteps
in $\omega_{ce}$ units as we vary the mass ratio, as detailed in
the second column of the table. The wall-clock time (WCT) for 20 timesteps,
the corresponding number of nonlinear iterations (NLI) for a nonlinear
tolerance of $10^{-10}$, and the estimated speedup with respect to
the $\Delta t=0.2\omega_{ce}^{-1}$ case are listed. The speedup is
estimated for the same final time in $\omega_{pi}$ units, assuming
the cost per timestep of the $\Delta t=0.2\omega_{ce}^{-1}$ run does
not fundamentally change over the span of the longer runs.\label{table-m2stream}}
\begin{tabular}{|c|c|c|c|c|}
\hline 
$m_{i}/m_{e}$ & $\omega_{ce}\Delta t$ & WCT (s) & NLI & Speedup\tabularnewline
\hline 
\hline 
1000 & 18 & 5.5 & 5.9 & 40\tabularnewline
\hline 
2000 & 36 & 5.5 & 5.9 & 80\tabularnewline
\hline 
5000 & 90 & 5.18 & 5.9 & 212\tabularnewline
\hline 
10000 & 180 & 5.09 & 5.9 & 430\tabularnewline
\hline 
\end{tabular}
\end{table}

We look next at the performance of the solver as a function of the
ion-electron mass ratio (which determines electron magnetization).
Table \ref{table-m2stream} shows that both the wall-clock time and
the number of nonlinear iterations stay constant as the mass ratio
increases from 1000 to 10000, i.e., by a factor of 10 (with other
numerical parameters unchanged), demonstrating complete insensitivity
of the performance of the algorithm to the mass ratio (i.e., to the
electron magnetization level) for this problem, and significant speedups
as the mass ratio increases, up to 430 for the 10000 mass-ratio case.

\subsection{Drift-wave instability}

\begin{figure}
\centering{}\includegraphics[width=0.4\columnwidth]{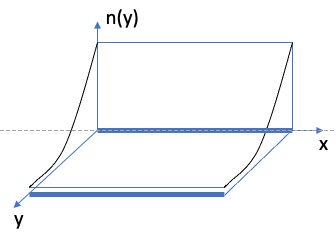}\caption{Density profile for the drift wave instability. The 2D domain is periodic
in $x$ and confined by perfect conductors in $y$. \label{fig:drift-wave-setup}}
\end{figure}

Finally, we consider a drift-wave instability driven by a non-uniform
plasma density profile in a 2D bounded slab of magnetized plasma.
The magnetic field is homogeneous and nearly perpendicular to the
density gradient. The drift waves simulated are low-frequency electrostatic
waves well below the ion cyclotron frequency and propagating almost
perpendicularly to the magnetic field, justifying the 2D simulation.
We choose a 2D simulation domain (with $L_{x}\times L_{y}=16\times32\lambda_{De}^{2}=32\times64\lambda_{Di}^{2}$)
with a mesh of $32\times64$ grid points. We set the $x$-direction
to be periodic and the $y$-direction to be bounded with a perfect
conductor. The magnetic field has a small component in the $x$-direction,
i.e., $B_{x}=0.01047\ll B_{z}=0.399986$, and $B_{y}=0$. The density
gradient is set to be $n(y)=\kappa L_{y}e^{-\kappa y}/(1-e^{-\kappa L_{y}})$
with $\kappa=0.14\,\lambda_{De}^{-1}=0.07\,\lambda_{Di}^{-1}$. Other
parameters are $m_{e}=1/25$, $m_{i}=1$, $v_{the}=10$, $v_{thi}=1$,
and $\omega_{ci}/\omega_{pi}=0.4$. These parameters are chosen identically
to those used in Ref. \citep{lee1976anomalous} (although with a different
normalization convention, as we employ ion units here). We employ
8 particles per cell and $\Delta t=0.04\,\omega_{pi}^{-1}=0.4\,\omega_{ce}^{-1}$,
the same as in Ref. \citep{lee1976anomalous}. We will also show results
with a larger timestep, $\Delta t=0.2\,\omega_{pi}^{-1}=2.0\,\omega_{ce}^{-1}$,
for comparison. A sketch of the plasma density profile in the 2D domain
is shown in Fig. \ref{fig:drift-wave-setup}.

\begin{figure}
\centering{}\includegraphics[width=1\columnwidth]{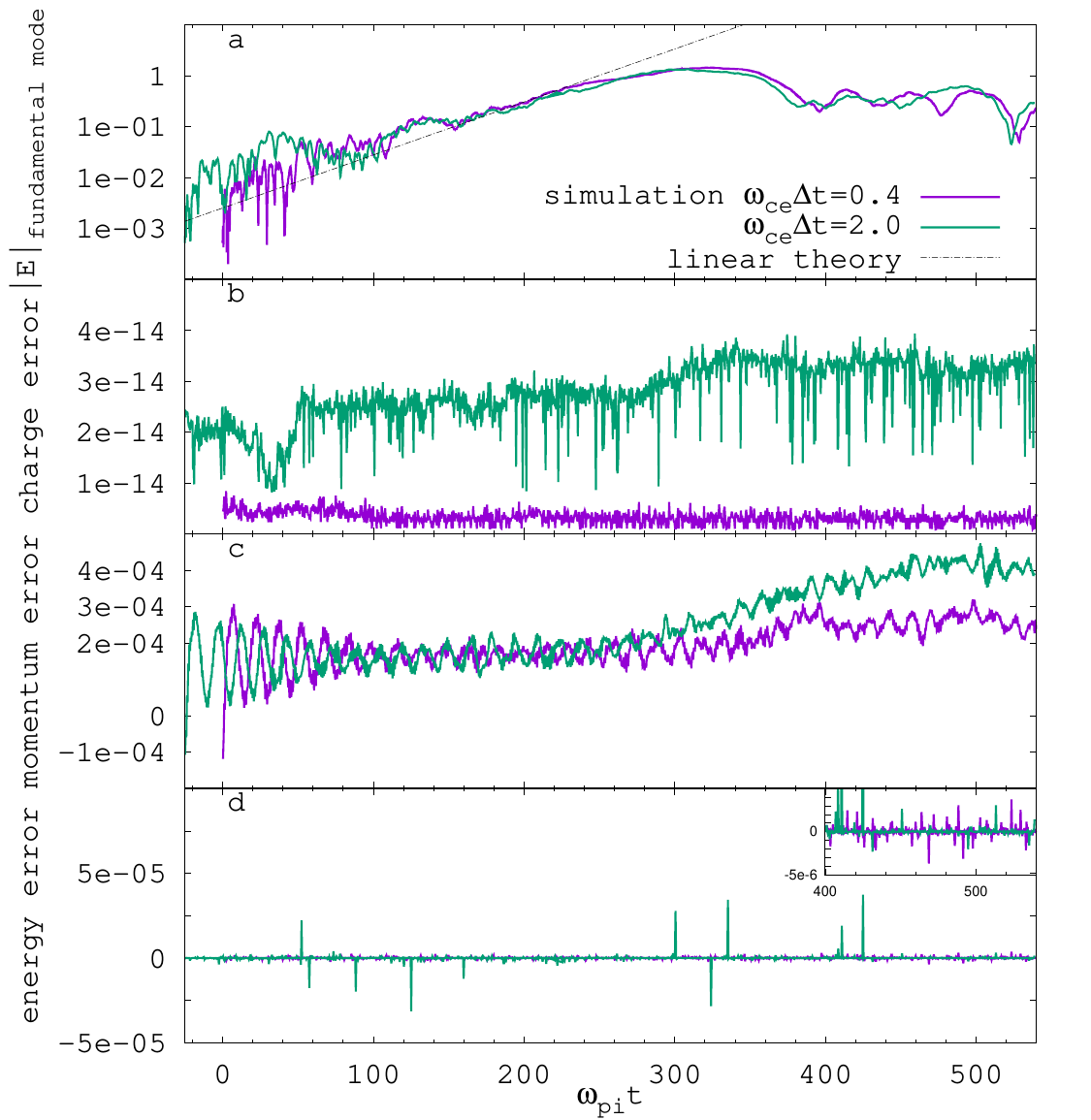}\caption{Simulation results for the drift-wave instability problem. Panel a
depicts the growth in the electric-field energy with two different
timesteps, $\Delta t=0.4\,\omega_{ce}^{-1},\,2.0\,\omega_{ce}^{-1}$.
Panels b, c, d depict time histories of the root-mean-square (rms)
of the charge continuity equation, the total momentum error (normalized
to $\sum_{\alpha}\sum_{p}m_{p}v_{th\alpha}$), and total energy error,
respectively (measured as in Fig. \ref{fig:Diocotron}). The $\Delta t=2.0\,\omega_{ce}^{-1}$
curves are shifted left in all panels by 25 time units to provide
a better comparison of the growing phase of the instability with the
$\Delta t=0.4\,\omega_{ce}^{-1}$ results. \textcolor{red}{The top-right
corner of panel d provides additional details of the energy error
over the time range {[}400-540{]}.} \label{fig:driftwave}}
\end{figure}
\begin{figure}
\centering{}\includegraphics[height=0.9\textheight]{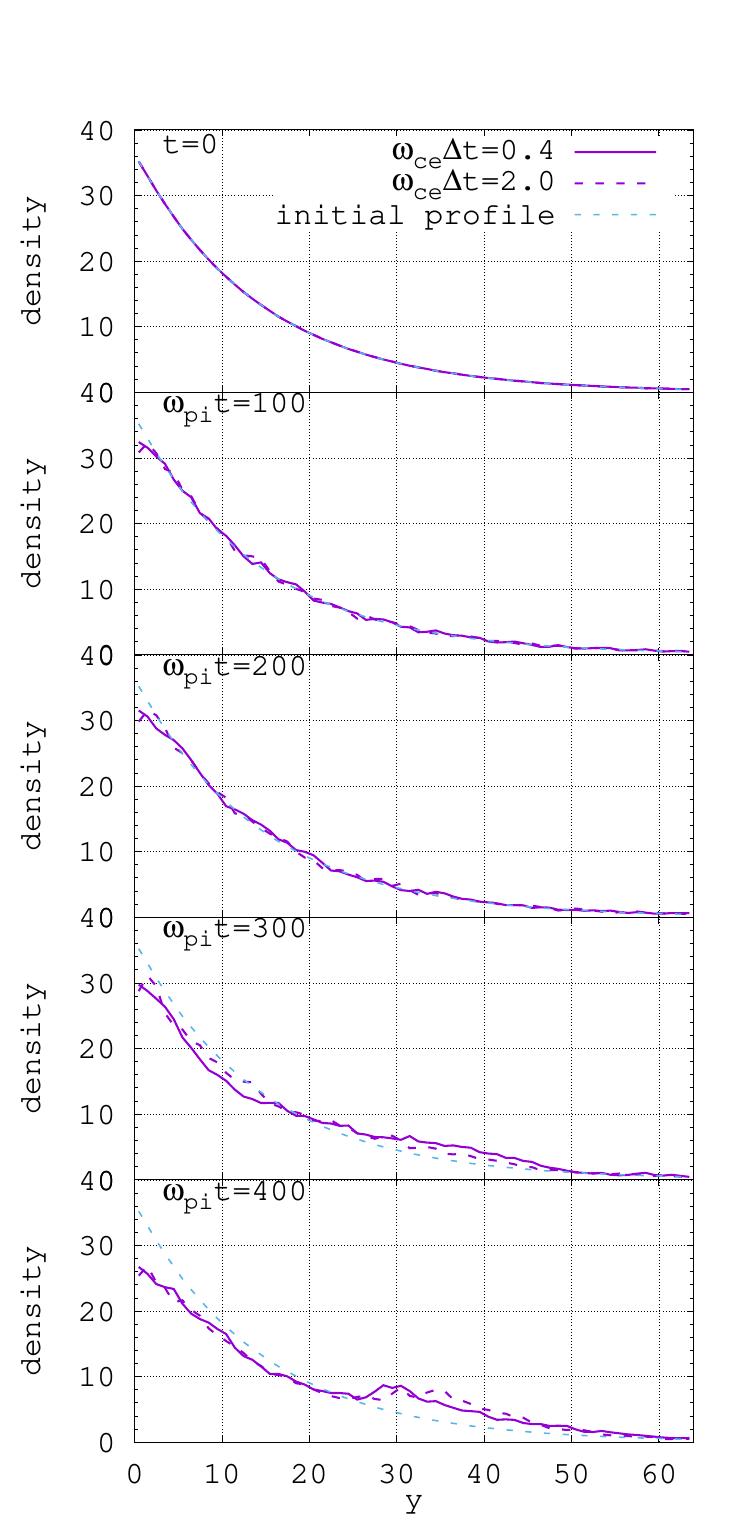}\caption{Ion density profiles at different times during the drift-wave instability
evolution. The density is averaged over the $x$ domain. The initial
density is an exponential profile. Four snapshots are depicted to
show the evolution of the density profile in $y$.\label{fig:density-drift-wave}}
\end{figure}

Simulation results are shown in Fig. \ref{fig:driftwave}, where we
find good agreement of the growth-rate of the fundamental mode with
the linear theory. The growth rate is obtained as the imaginary part
of the solution to the linear dispersion relation \citep{lee1976anomalous},
\begin{equation}
\sum_{\alpha}\frac{1}{T_{\alpha}}\left[1+\frac{\omega+\omega_{\alpha}^{\star}}{\sqrt{2}k_{\parallel}v_{th\alpha}}I_{0}(\hat{b}_{\alpha})e^{-\hat{b}_{\alpha}}Z\left(\frac{\omega}{\sqrt{2}k_{\parallel}v_{th\alpha}}\right)\right]=0,
\end{equation}
where $\alpha$ denotes species, $v_{th\alpha}=\sqrt{T_{\alpha}/m_{\alpha}}$,
$k_{\parallel}=k_{y}\cos\theta$ with $\theta$ the angle between
$\mathbf{B}$ and $\mathbf{k}$, $I_{0}$ is the Bessel function of
the first kind and order 0 \citep{abramowitz1988handbook}, $\hat{b}_{\alpha}=r_{\alpha}^{2}(k_{x}^{2}+k_{y}^{2})$,
$r_{\alpha}=v_{th\alpha}/\omega_{c\alpha}$, $\omega^{\star}=k_{y}T_{e}\kappa/m_{i}\omega_{ci}$,
$\omega_{e}^{\star}=-\omega^{\star}$, $\omega_{i}^{\star}=\omega^{\star}T_{i}/T_{e}$,
and $Z$ is the plasma dispersion function \citep{fried2015plasma}.
In this simulation, $k_{x}r_{i}=0.12$ and $k_{y}r_{i}=0.49$ for
the longest wavenumber mode. The instability evolves fairly slowly.
It takes a relatively long time (\textasciitilde 250 $\omega_{pi}^{-1}$)
for the mode to grow exponentially (with a growth rate of $\gamma=0.024\omega_{pi}$),
and another \textasciitilde 50 $\omega_{pi}^{-1}$ to saturate nonlinearly.
Nonlinear saturation occurs due to the flattening of the density profile.
In the other panels of Fig. \ref{fig:driftwave}, we again see excellent
charge and energy conservation, and reasonable momentum conservation.
Figure \ref{fig:density-drift-wave} shows a few snapshots in time
of the ion density profile. It is clear that the profile is able to
maintain the initial exponential profile up to $t\sim230\omega_{pi}^{-1}$
in the linear stage, after which the density starts to flatten out
and the instability transitions into the nonlinear stage.

A few comments are in order when comparing against the results in
Ref. \citep{lee1976anomalous}. We firstly note that the simulation
algorithms are quite different. The reference reports using a guiding-center
model simulated by a dipole-expansion technique with finite-size particles
and explicit time-integration, while we use an asymptotic-preserving
implicit PIC algorithm. Nevertheless, our simulations find good agreement
with the linear theory for about as long as the simulations in the
reference. Both sets of results run into a slower-growing phase before
getting into the saturated nonlinear stage, which is caused by the
flattening of the density gradient. While it appears that the instability
gets into clear exponential growth earlier in the simulations in the
reference than in ours, we speculate that this is likely due to the
initialization procedure.

Another obvious difference between the two approaches is the treatment
of particle boundary conditions. As is well-known, the treatment of
particle boundaries in the guiding-center model can be troublesome
\citep{lee1978simulation}, whereas in our algorithm the boundary
treatment is quite natural and straightforward, as we can seamlessly
transition particles to full-orbit integration if they hit the wall.
In our simulations, we do this by estimating the sub-timestep according
to the full-orbit time estimator described in Sec. \ref{subsec:Boundary-treatment}
if a particle is reflected by the wall. We have found that this simple
treatment affords both long-term accuracy and robustness in practice.

There are significant differences between $\Delta t=0.4\,\omega_{ce}^{-1}$
and $\Delta t=2.0\,\omega_{ce}^{-1}$ in terms of performance. For
instance, for 20 timesteps, the average number of nonlinear iterations
per timestep is 4.9 and 8.1, respectively, with corresponding wall-clock
times of 110 s and 182 s. We see that, as the timestep is increased
by a factor of 5, the average number of Newton iterations increases
only by a factor of about 1.6, and the CPU time increases proportionally,
resulting in a speedup of about 3.0 to reach the same final simulation
time (in $\omega_{pi}$ units). It is also worth noting that, for
$\Delta t=2.0\,\omega_{ce}^{-1}$, electrons may travel on average
a distance of $v_{the}\Delta t=2\lambda_{Di}$, about 14$\%$ of the
gradient length scale ($1/\kappa\simeq14\,\lambda_{Di}$). For larger
timesteps, it may be necessary to include the gradient length scale
in the timestep estimator, as proposed in Ref. \citep{ricketson2020energy}.

\begin{table}
\centering{}\caption{Solver performance with respect to mass ratio $m_{i}/m_{e}$ for the
drift-wave instability problem. All runs employ a fixed timestep \textcolor{red}{${\color{black}{\normalcolor \Delta t=0.04}\,\omega_{pi}^{-1}}$},
corresponding to different timesteps in $\omega_{ce}$ units as we
vary the mass ratio, as detailed in the second column of the table.
The wall-clock time (WCT) for 20 timesteps, the corresponding number
of nonlinear iterations (NLI), and the estimated speedup with respect
to the $\Delta t=0.4\,\omega_{ce}^{-1}$ case are listed.\label{table-drift-wave}}
\begin{tabular}{|c|c|c|c|c|}
\hline 
$m_{i}/m_{e}$ & $\omega_{ce}\Delta t$ & WCT (s) & NLI & Speedup\tabularnewline
\hline 
\hline 
25 & 0.4 & 110 & 4.9 & 1\tabularnewline
\hline 
100 & 1.6 & 113 & 5.3 & 4.8\tabularnewline
\hline 
1000 & 16 & 133 & 6.4 & 41.3\tabularnewline
\hline 
2500 & 40 & 182 & 8.1 & 60.4\tabularnewline
\hline 
\end{tabular}
\end{table}
It is again of interest to examine the performance of the algorithm
as we increase the ion-electron mass ratio for the drift-wave problem.
Table \ref{table-drift-wave} shows the CPU time and nonlinear iteration
count when varying the mass ratio from 25 to 2500 for $\Delta t=0.04\,\omega_{pi}^{-1}$
\textcolor{black}{while keeping all other numerical parameters fixed.
It shows that the wall-clock time only increases by about a factor
of 1.6 as the mass ratio increases by a factor of 100, again demonstrating
significant insensitivity to the electron magnetization, as expected
from an AP formulation, and the potential for significant speedups
as the mass ratio increases.}

\section{Discussion and summary}

\label{sec:conclusions}

In this study, we have demonstrated an implicit, conservative PIC
algorithm capable of stepping over gyration timescales in strongly
magnetized regimes without apparent loss of accuracy and while conserving
charge and energy exactly. The approach allows adjustment of the particle
orbit timestep to the presence of boundaries, thus allowing a straightforward
boundary treatment. Our implementation is for the time being electrostatic
(which avoids complexities due to the presence of magnetization currents
in the electromagnetic case \citep{brizard2016variational}), and
considers a uniform magnetic field (which avoids the $\nabla B$ drift,
and allows the use of a straightforward Crank-Nicolson integrator).
Even in this reduced context, the results in this study are remarkable
in that they demonstrate the ability of the AP PIC algorithm to produce
orders-of-magnitude computational savings in strongly magnetized environments
vs. the well-resolved subcycled method, without spoiling conservation
properties or long-term accuracy. Key to the development of the method
is the break-up of the particle orbit into substeps in which the particle
velocity is considered constant. Substeps are determined so as to
resolve electric-field length scales. These substeps may traverse
an arbitrary number of cells, with the local charge and energy deposition
to the crossed cells computed \emph{a posteriori}. If a particle hits
a wall, we employ a simple second-order timestep estimator and apply
a full-orbit treatment for that particle for that timestep. As a result
of decoupling particle cell crossings (required for strict charge
and energy conservation) from the actual orbit integration, orders
of magnitude CPU speedups are realized (speedups of more than two
orders of magnitude have been demonstrated for results of equivalent
accuracy). It is important to note that the field solver is able to
step over gyro-motion timescales (and associated Bernstein waves)
without requiring the particles to resolve their gyro-motion.

Future work will consider several directions. In the short term, we
will\textcolor{black}{{} extend the CN integrator to include $\nabla B$-drift
corrections (as proposed in Ref. \citep{ricketson2020energy}) to
allow for nontrivial magnetic field configurations. In this case,
the particle pushing can be done in a similar fashion as in this study,
that is, a particle will be pushed in straight lines within sub-timesteps,
with additional average forces and sub-timestep constraints related
to the magnetic field gradient. Next, we will extend the approach
to the electromagnetic regime in which the magnetic field is determined
self-consistently. In this case, the effects of magnetization current
need to be included. Longer-term, we will consider extending the AP
orbit integrator to include finite-Larmor-radius effects. The challenge
there will be to generalize the mover in Ref. \citep{ricketson2020energy}
to the gyro-kinetic regime (for which $k\rho_{c}\sim\mathcal{O}(1)$,
with $\rho_{c}$ the gyroradius and $k$ the characteristic inverse
length scale), and incorporate it without losing charge and energy
conservation in the implicit PIC framework.}

\section*{Acknowledgments}

The authors acknowledge useful conversations with Lee Ricketson. This
research has been funded by the Applied Mathematics Research program
of the Department of Energy Office of Applied Scientific Computing
Research, used computing resources provided by the Los Alamos National
Laboratory Institutional Computing Program, and was performed under
the auspices of the National Nuclear Security Administration of the
U.S. Department of Energy at Los Alamos National Laboratory, managed
by Triad National Security, LLC under contract 89233218CNA000001.

\bibliographystyle{ieeetr}
\bibliography{dpicLargeDt}

\end{document}